\documentclass[prb,preprint,endfloats]{revtex4}

\usepackage{graphicx}
\begin{document}

\title{Magnetoelectrical control of nonreciprocal microwave response in a multiferroic helimagnet}

\author{Y. Iguchi, Y. Nii, and Y. Onose} 
\affiliation{Department of Basic Science, University of Tokyo, Tokyo, 153-8902, Japan}


\begin{abstract}
\bf{
Control of physical property in terms of external fields is essential for contemporary technologies. The conductance can be controlled by a gate electric field in a field effect transistor, which is a main component of the integrated circuit. Optical phenomena induced by an electric field such as electroluminescence and electrochromism are useful for display and other technologies. Control of microwave propagation seems also imperative for future wireless communication technology. Microwave properties in solids are dominated mostly by magnetic excitations, which cannot be easily controlled by an electric field. One of the solutions for this problem is utilizing magnetically induced ferroelectrics (multiferroics). Here we show that microwave nonreciprocity, which is difference between oppositely propagating microwaves, can be reversed by the external electric field in a multiferroic helimagnet Ba$_2$Mg$_2$Fe$_{12}$O$_{22}$. This result offers a new avenue for the electrical control of microwave properties.

}
\end{abstract}
\maketitle

Multiferroics have been investigated intensively since the discovery of magnetically induced ferroelectrics in perovskite manganites\cite{KimuraNature}. In this class of materials, there exists giant magnetoelectric effect, which is the electric-field-induced change of magnetization and its reciprocal effect, namely the magnetic-field-induced change of electric polarization. The magnetoelectric coupling is also valid in the course of magnetic oscillation. The spin wave excitation is coupled to an alternating electric field as well as an alternating magnetic field in multiferroics\cite{PimenovNatPhys2006,KatsuraPRL2007}. The dynamical magnetoelectric coupling gives rise to unique nature of electromagnetic wave. The refractive indices of oppositely propagating electromagnetic waves are different from each other, which is denoted as nonreciprocal directional dichroism. This is certainly realized in a transverse conical magnetic state, one of the prototypical spin arrangements exhibiting multiferroic properties, as shown in Fig. 1a. This magnetic state has net magnetization $\textbf{M}$ along one direction. The magnetic moment components perpendicular to $\textbf{M}$ rotate along the conical wave vector $q_0$ perpendicular to $\textbf{M}$. According to the spin current mechanism\cite{KatsuraPRL2005}, the rotating components give rise to static electric polarization $\textbf{P}$. Thus $P$ as well as $M$ may be oscillated in the spin excitations of transverse conical state. In particular one of the Nambu-Goldstone magnons ($q=q_0$ excitation) simultaneously induces alternating $P$ ($\Delta P$) and alternating $M$ ($\Delta M$)\cite{Miyahara2014}. Therefore this mode is electrically and magnetically active and shows strong magnetoelectric effect. As a result the large magnetoelectric coupling induces the term of $\textbf{k}\cdot(\textbf{P}\times\textbf{M})$ in the refractive index of electromagnetic wave indicating the nonreciprocal directional dichroism that can be controlled by a dc electric field as well as a dc magnetic field. Recently the controllable optical nonreciprocity has been observed for a spin excitation in terahertz regime at large magnetic field $3\leq \mu_0H \leq7$\ T\cite{Takahashi2011}, which is ascribed to $q=q_0$ excitation. Because this mode is one of Nambu-Goldstone modes, the frequency should be much lower in the absence of magnetic field. Here we report the controllable nonreciprocity in ubiquitous frequency ($10-15$ GHz) and magnetic field (160 mT) ranges in a multiferroic helimagnet Ba$_2$Mg$_2$Fe$_{12}$O$_{22}$.  While microwave nonreciprocity was previously observed in a chiral magnet, it depends on the crystal chirality and cannot be reversed by an electric field\cite{Okamura2013}. In Ba$_2$Mg$_2$Fe$_{12}$O$_{22}$ the electric polarization can be controlled by a low magnetic field\cite{IshiwataScience2008}. Related materials show magnetoelectric response above room temperature\cite{Kimura2012}. The microwave functionality in this novel material may contribute to the future technology.

\begin{figure}
\begin{center}
\includegraphics*[width=13cm]{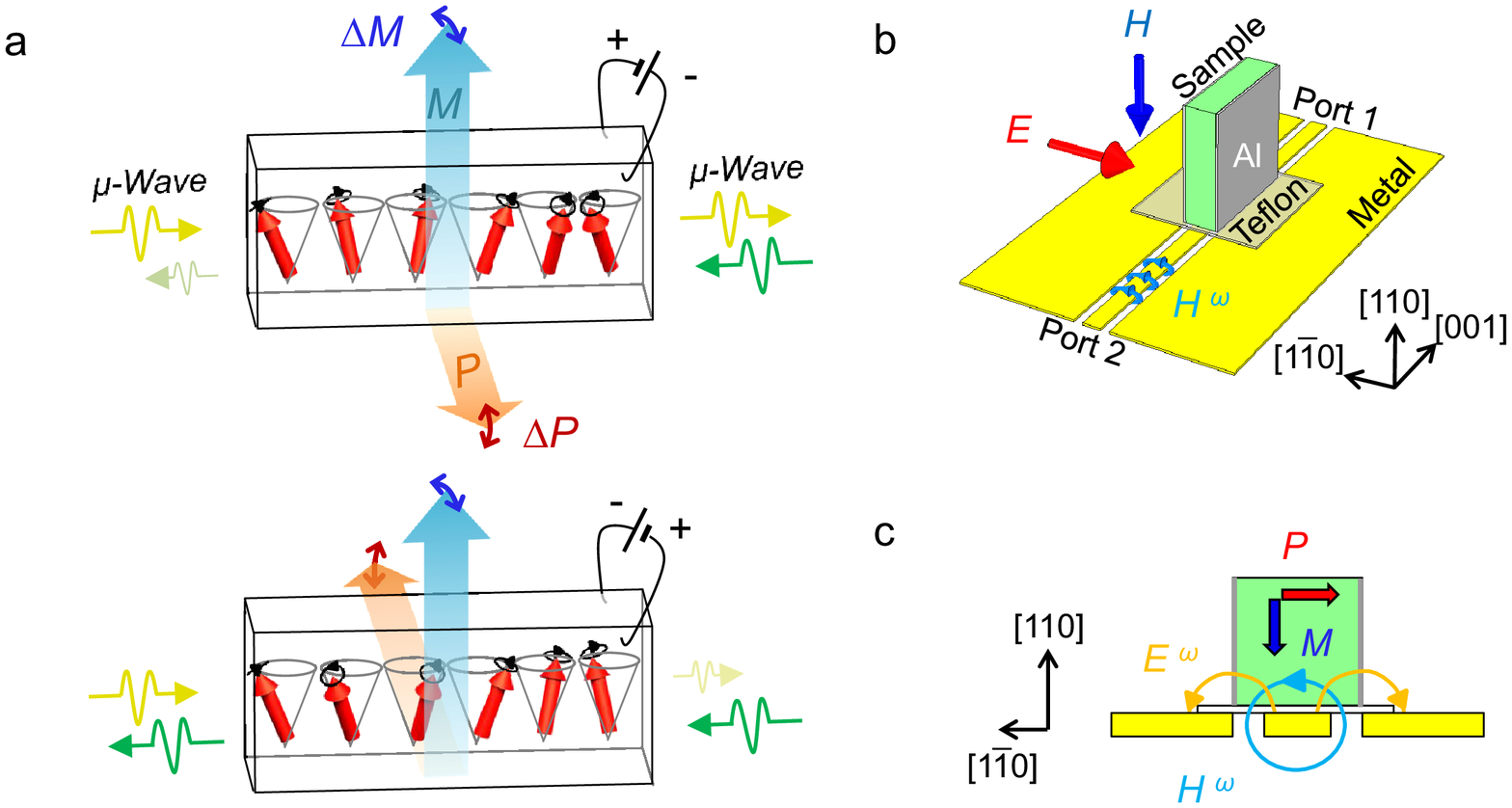}
\caption{\textbf{Electrical control of nonreciprocal microwave transmission.}
(\textbf{a})\ Sketch of the nonreciprocal microwave transmission in ferroelectric helimagnets. The strong magnetoelectric coupling of $q = q_0$ Nambu-Goldstone magnon in the transverse conical magnetic state provides the nonreciprocity, which is the difference of refractive indices between oppositely propagating microwaves. The microwave nonreciprocity as well as spin helicity can be controlled by the electric field. (\textbf{b})\ Experimental setup of the microwave measurement. (\textbf{c})\ Illustration of the cross-section view of the (001) plane and the alternating magnetic field $H^{\omega}$ and electric field $E^{\omega}$ of microwave in the experimental setup.}
\end{center}
\end{figure}

\section*{Results}
\subsection*{Magnetostructural variation in magnetic field}

The left panel of Fig. 2a shows the crystal structure of Ba$_2$Mg$_2$Fe$_{12}$O$_{22}$. The crystal structure is composed of alternately stacking two blocks denoted as S and L blocks. Within each block the magnetic moments are ferrimagnetically ordered. As a result L and S blocks have large and small net magnetic moments, respectively. Figure 2b shows the magnetization curve at 6 K for Ba$_2$Mg$_2$Fe$_{12}$O$_{22}$. At low temperature, the magnetic moments of S and L blocks show conical magnetic orderings where the wave vector of spin structure is along the [001] axis\cite{Momozawa1986}. Therefore there is a small but finite spontaneous magnetization. When the magnetic field is increased, the magnetization curve shows several kinks that reflect magnetostructural transitions as indicated by inverted triangles in Fig. 2b. While the helical plane is perpendicular to the [001] axis at zero magnetic field, the helical plane is slanted and ferroelectricity is induced parallel to the [1$\bar{1}$0] axis in a small magnetic field along the [110] axis (FE1). When the magnetic field is increased further, the helical plane becomes perpendicular to the [110] axis and the transverse conical state is realized. The spiral wave vector is $(0,0,3/4)$ between 60 mT and 200 mT (FE2), and $(0,0,3/2)$ between 200 mT and 4.5 T (FE3)\cite{IshiwataPRB2010}. The magnetic structure in FE2 state is shown in the right panel of Fig. 2a. Above 4.5 T the collinear ferrimagnetic state shows up and the ferroelectricity is quenched (PE).

\begin{figure}
\begin{center}
\includegraphics*[width=12cm]{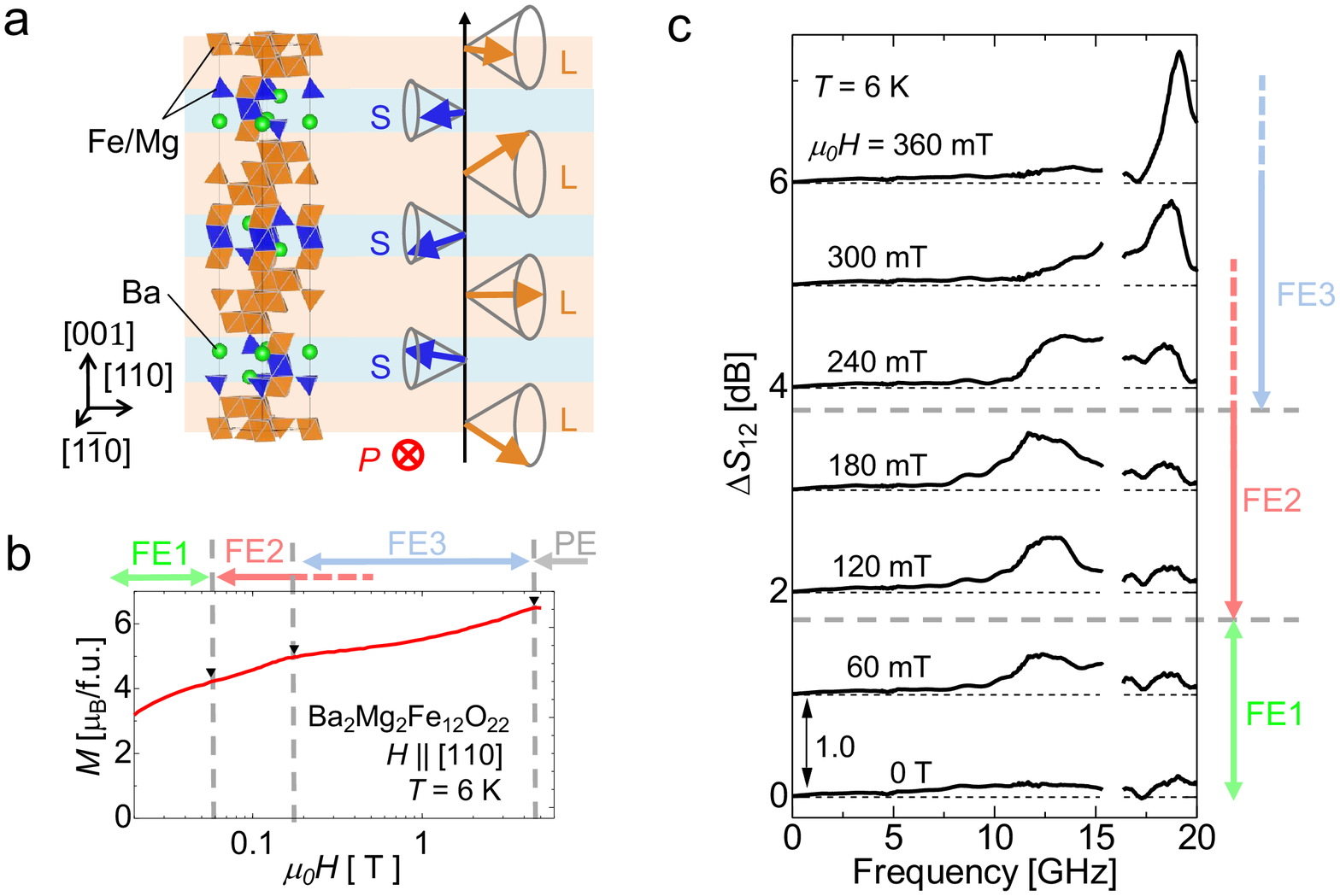}
\caption{\textbf{Crystal structure and magnetic properties of Ba$_2$Mg$_2$Fe$_{12}$O$_{22}$.} (\textbf{a})\ Crystal structure of Ba$_2$Mg$_2$Fe$_{12}$O$_{22}$ and spin structure in FE2 state, in which we demonstrate the control of microwave nonreciprocity. (\textbf{b})\ Magnetization curve of Ba$_2$Mg$_2$Fe$_{12}$O$_{22}$ at 6K in the magnetic field parallel to the [110] direction. (\textbf{c})\ Microwave absorption $\Delta S_{12}$ in several magnetic fields at 6 K. Data between 15.3 GHz and 16.3 GHz are omitted because of strong noise in this frequency region.}
\end{center}
\end{figure}

\subsection*{Microwave absorption}
Figure 2c shows $\Delta S_{12}$ spectra at various magnetic fields at 6 K that are measured in the experimental setup shown in Fig. 1b. The $\Delta S_{12}$ reflects the absorption of microwave owing to magnetic resonance in the course of propagation from port 2 to port 1. For the detail of experimental setup and precise definition of $\Delta S_{12}$, see Methods. In the FE1 state, a small and broad absorption peak is observed around 12 GHz. Stronger microwave absorption appears in the same frequency range in the FE2 state. In the FE3 state, the peak shape and frequency become sharper and larger as the magnetic field is increased. We demonstrate the control of nonreciprocity in the FE2 state because the measurement sensitivity in the frequency range of $9-14$\ GHz is better than that of higher frequency region in our experimental setup.

\subsection*{Magnetoelectric control of nonreciprocity}
Figure 3 demonstrates the magnetoelectrical control of nonreciprocal microwave absorption. $\Delta S_{21}$ is microwave absorption spectrum similar to $\Delta S_{12}$ but the microwave progataion direction is opposite. We perform the poling procedure with use of external electric field in order to fix the spin helicity (for the detail, see Methods). After the poling procedure we turn off the external electric field and change the magnetic field to certain value and then measure the $\Delta S_{12}$ and $\Delta S_{21}$. Figures 3a and 3c show the spectra measured after the poling procedure with the electric field $E$ of $+$0.5 MVm$^{-1}$, and Figs. 3b and 3d with $E = -$0.5 MVm$^{-1}$. The measured magnetic field $H$ is $+$160 mT for Figs. 3a and 3b, and $-$160 mT for Figs. 3c and 3d. For all the cases there are a clear differences between $\Delta S_{12}$ and $\Delta S_{21}$ which indicate the microwave nonreciprocity. The nonreciprocity is reversed by the inversion of either $E$ or $H$, but it is unchanged by the simultaneous inversion of $E$ and $H$. 

\begin{figure}
\begin{center}
\includegraphics*[height=8cm]{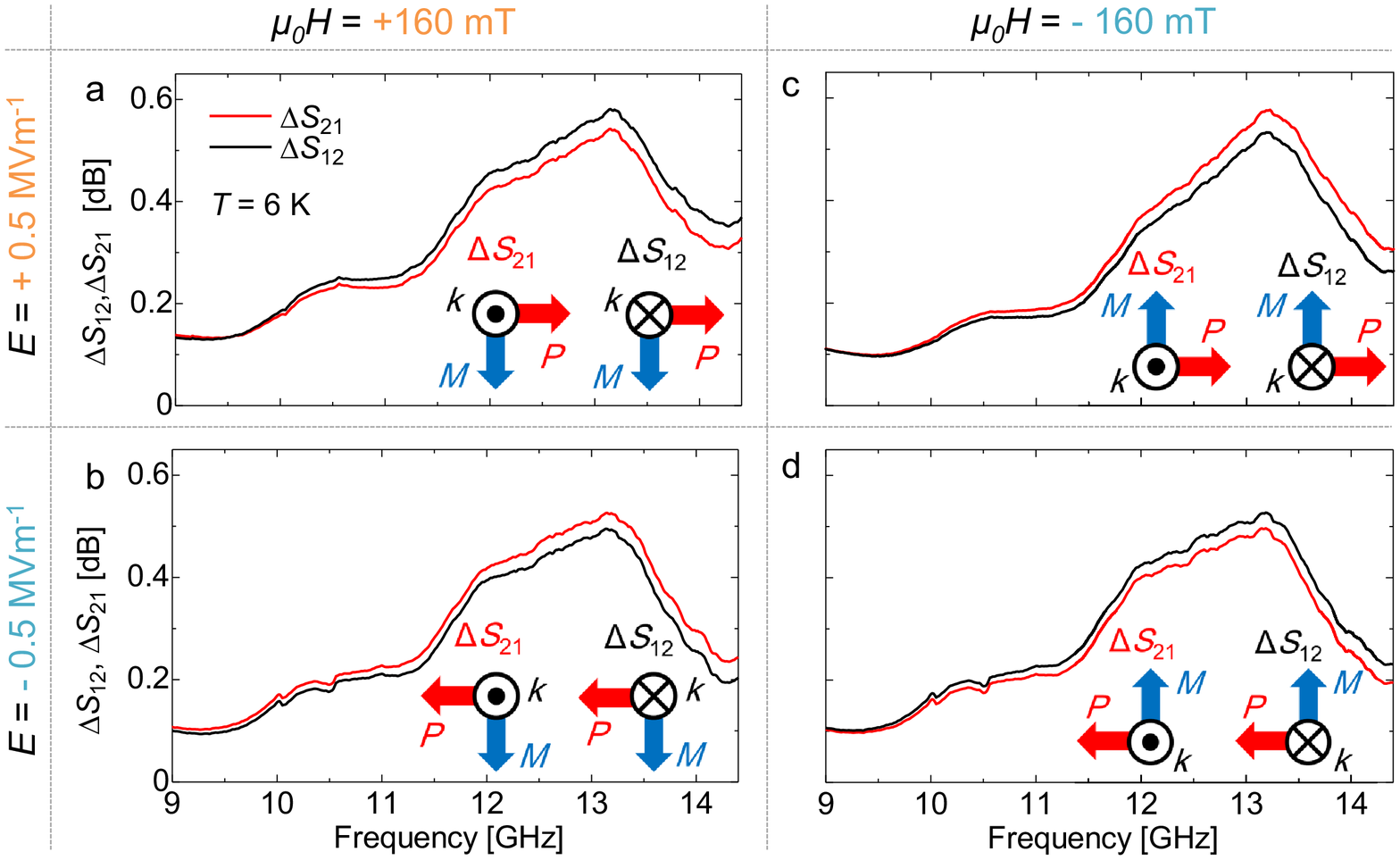}
\caption{\textbf{Magnetoelectrical control of microwave nonreciprocity.}  Microwave absorption $\Delta S_{12}, \Delta S_{21}$ at 6 K in $\pm$160 mT. The poling fields are $E = \pm$0.5 MVm$^{-1}$. Inset illustrates the directional relation of $P$, $M$, and $k$ for $\Delta S_{12}$ and $\Delta S_{21}$ measurements.}
\end{center}
\end{figure}

In order to further study the effect of external field we present the poling electric fields dependence of microwave nonreciprocity $\Delta S_{12} - \Delta S_{21}$ in Fig. 4a. The sign of nonreciprocity reflects that of $E$, and the magnitude monotonically increases with $E$. In Fig. 4c we show the integrated intensity of nonreciprocity $I_{12}$ between 9 GHz and 14.4 GHz at $\mu_0 H = \pm$160 mT as a function of $E$. The sign of $I_{12}$ depends on that of $E$ and $H$, and the magnitude of $I_{12}$ gradually increases in the low $E$ region and tends to be saturated around 0.5 MVm$^{-1}$. Actually similar $E$ dependence is discerned in the polarization. Therefore these field dependences are dominated by the ferroelectric domain population.

\begin{figure}
\begin{center}
\includegraphics*[width=15cm]{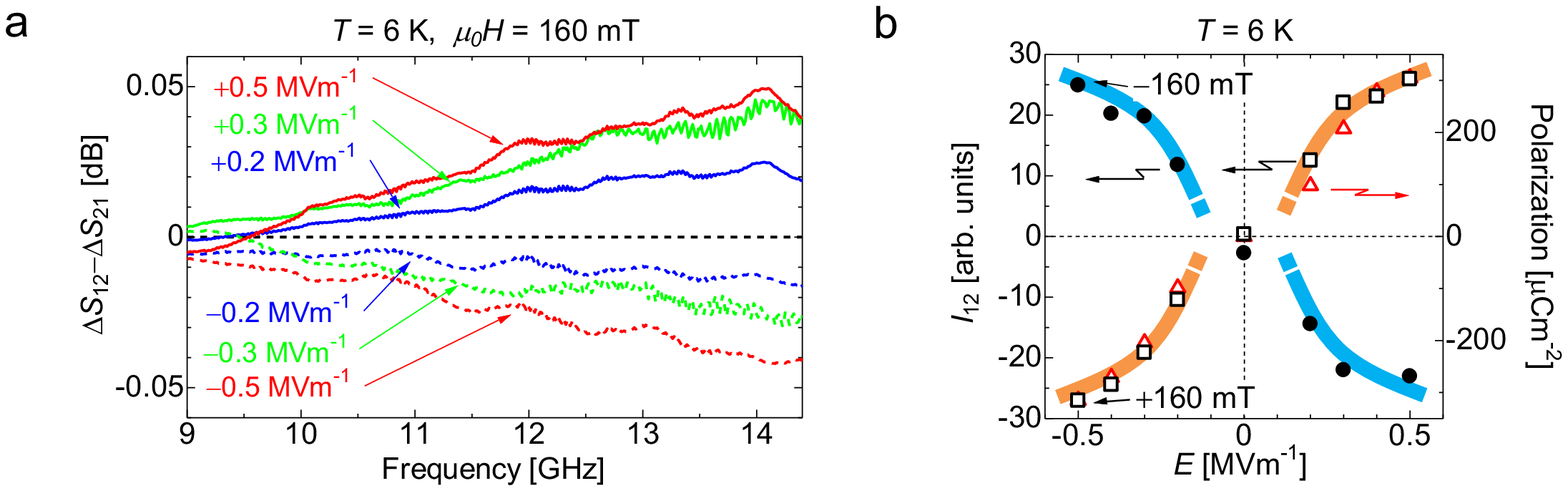}
\caption{\textbf{Electric field dependence of microwave nonreciprocity.} (\textbf{a})\ Nonreciprocities $\Delta S_{12} - \Delta S_{21}$ at $+$160 mT measured after various electric fields poling. (\textbf{b})\ Integrated intensities of nonreciprocity $I_{12}$ at $+$160 mT (open square) and $-$160 mT (closed circle) are plotted as a function of the poling electric field $E$. The polarization at $+$160 mT (open triangle) is also plotted as a function of $E$ for the comparison. The solid lines are merely guides to eyes.}
\end{center}
\end{figure}

\section*{Discussion}
We observed the controllable nonreciprocity for the lowest energy magnetic resonance mode in the transverse conical state (FE2). The observed nonreciprocity cannot be ascribed to the effect of magnetic dipole interaction, because it should not be changed by the poling electric field in the case of magnetic dipolar nonreciprocity. As explained in the introduction, the lowest energy magnetic resonance in transverse conical state is $q = q_0$ Nambu-Goldstone magnon mode that is expected to show the large nonreciprocity along to $\textbf{P}\times\textbf{M}$, which is quite consistent with the present observations. While similar magnon modes were previously observed at large magnetic fields in the THz region for other spiral magnets\cite{Takahashi2011,TakahashiPRL2013,TakahashiPRB2016}, here we have observed it at a low magnetic field in GHz region. The magnitude of nonreciprocity in this study is as large as $6-8$\% which is smaller than those observed in THz region. One of the reason for this difference is that the intensity of pure magnetic excitation becomes relatively large and therefore the relative nonreciprocity becomes small in this case of the spontaneous conical state. The advantage of microwave nonreciprocity is compatibility with other microwave technologies. For example, by utilizing the high-Q resonator the nonreciprocity can be adequately enhanced. While the electrical control of microwave property has been extensively investigated for multiferroic heterostructures mainly with use of mechanical strain-mediated magnetoelectric coupling\cite{Srinivasan2010}, the controllable nonreciprocity in the transverse conical state seems more useful. Thus the new microwave functionality of controllable nonreciprocity has large potential for practical application.

\section*{Methods}
\subsection*{Sample preparation and magnetization measurement}
Single crystals of Ba$_2$Mg$_2$Fe$_{12}$O$_{22}$ was grown by the flux method\cite{Momozawa1986}. The crystal orientations are determined by X-ray diffraction measurement. Magnetization curve was measured by using a superconducting quantum interference device magnetometer (Magnetic Property Measurement System, Quantum Design) at 6 K in magnetic fields parallel to the [110] direction. 

\subsection*{Polarization measurement and poling procedure}
Spontaneous electric polarization along the [1$\bar{1}$0] direction at 6 K was obtained by the integration of measured displacement currents. Before the current measurement we perform the following poling procedure to fix the spin helicity similarly to the previous study\cite{IshiwataScience2008}. After cooling down to 50 K without external fields, the electric field is applied to the [1$\bar{1}$0] direction and then the magnetic field as large as 5 T is applied parallel to the [110] direction. Next the magnetic field is decreased to 1 T followed by cooling down to 6 K. Finally the electric field is removed. The displacement current was measured with sweeping the magnetic field.

\subsection*{Microwave spectroscopy}
We fabricate the microwave device composed of the Ba$_2$Mg$_2$Fe$_{12}$O$_{22}$ sample and a microwave coplanar waveguide shown in Fig. 2c in order to measure the microwave response after electric poling procedure. The dimensions of the sample are $1.1\times1.1\times0.3$ mm$^3$. The largest plane is perpendicular to the [1$\bar{1}$0] direction and the two longer sides are parallel to the [110] and [001] directions. The aluminum electrodes are attached to the largest sample planes for electric field application. The coplanar waveguide is designed so that the characteristic impedance matches 50 $\Omega$. The width of the strip line is 0.2 mm and the gap between the strip line and the ground plane is 0.05 mm. The sample is put on the coplanar waveguide so that the [001] direction is parallel to the wave guide. A teflon sheet with the thickness of 20 $\mu$m is inserted between the sample and the waveguide for the insulation. The microwave response was measured in a superconducting magnet with use of a vector network analyzer (E5071C, Agilent). Before the microwave measurement we perform the poling procedure similarly to the polarization measurement. For the measurements at negative magnetic fields the magnetic field during the poling is also negative. The microwave absorption spectra $\Delta S_{12}$ at $H$ is defined as $S_{12}$(5 T) $-$ $S_{12}(H)$, where $S_{12}$ is the microwave transmittance from port 2 to port 1. Because the magnetic resonance frequency is enough over the measurement range at 5 T, the microwave absorption owing to magnetic resonance can be obtained by this formula. The microwave absorption with opposite wave vector $\Delta S_{21}$ is also obtained by a similar procedure.

\section*{Acknowledgements}
The authors thank S. Hirose for fruitful discussion. This work was in part supported by the Grant-in-Aid for Scientific Research(Grants No.25247058, No.16H04008, No.15K21622). Y. I. is supported by JPSJ fellows(No.16J10076).

\section*{Author contributions}
Y. I. carried out crystal growth, pyroelectric current measurement, coplanar waveguide preparation, microwave measurement, and analyzed data. Y. N. contributed to the modification of microwave experimental set up for the application of electric field. Y. O. supervised the project. Y. I. wrote the paper through the discussion and assistance from Y. N. and Y.O.

\section*{Additional information} 
Correspondence and requests for materials should be addressed to Y. I. (yiguchi@g.ecc.u-tokyo.ac.jp)

\newpage

\end{document}